\documentclass[%
 reprint,
 amsmath,amssymb,
 aps,
floatfix,
]{revtex4-2}
\usepackage{graphicx}
\usepackage{dcolumn}
\usepackage{bm}
\usepackage{subfigure}
\usepackage[colorlinks,linkcolor=blue,citecolor=blue,urlcolor=blue,hyperindex,breaklinks]{hyperref}
\allowdisplaybreaks[4]
\usepackage{subfigure}
\usepackage{xcolor}
\usepackage{soul}
\soulregister\cite7
\soulregister\ref7

\begin{document}
\title{Quantum heat diode versus light emission in circuit quantum electrodynamical system}
\author{Yu-qiang Liu$^1$, Yi-jia Yang$^1$ and Chang-shui Yu$^{1,2}$}
\email{Electronic address: ycs@dlut.edu.cn}
\affiliation{$^1$School of Physics, Dalian University of Technology, Dalian 116024,
P.R. China}
\affiliation{$^2$DUT-BSU Joint Institute, Dalian University of Technology, Dalian, 116024, China}

\date{\today}

\begin{abstract}
Precisely controlling heat transfer in a quantum mechanical system is particularly significant for designing quantum thermodynamical devices.  With the technology of experiment advances, circuit quantum electrodynamics (circuit QED) has become a promising system due to controllable light-matter interactions as well as flexible coupling strengths.  In this paper, we design a thermal diode in terms of the two-photon Rabi model of the circuit QED system. We find that the thermal diode can not only be realized in the resonant coupling but also achieve better performance, especially for the detuned qubit-photon ultrastrong coupling. We also study the photonic detection rates and their nonreciprocity, which indicates similar behaviors with the nonreciprocal heat transport.  This provides the potential to understand thermal diode behavior from the quantum optical perspective and could shed new insight into the relevant research on thermodynamical devices.
\end{abstract}
\maketitle
\section{Introduction}
Circuit quantum electrodynamics  \cite{devoret2007circuit, RevModPhys.93.025005}, which studies and controls light-matter interaction at the quantum level has become a fascinating topic for quantum optics, quantum thermodynamics, and condensed matter physics. For example, it has gradually been a toolbox of quantum optics on-chip \cite{devoret2007circuit, RevModPhys.93.025005, niemczyk2010circuit, PhysRevA.82.043804}. In particular, the circuit QED has also provided a prominent platform to precisely control and manipulate the heat to design quantum thermal machines \cite{tan2017quantum, PhysRevApplied.6.054014, PhysRevB.94.235420, karimi2017coupled, senior2020heat, PhysRevApplied.15.054050} and leads to the development of quantum thermodynamics \cite{doi:10.1080/00107514.2016.1201896, kurizki2022thermodynamics}. 
Concerning the latter, substantial efforts in recent years have been devoted to studying various quantum thermal machines in order to achieve some particular functions such as refrigerators and heat engines \cite{PhysRevLett.108.070604, PhysRevE.85.061126, PhysRevE.90.052142, PhysRevB.93.041418,PhysRevB.94.184503,tan2017quantum, PhysRevApplied.6.054014,PhysRevB.94.235420,PhysRevB.93.041418,PhysRevB.94.184503, binder2018thermodynamics}, heat switches \cite{karimi2017coupled}, thermal diodes \cite{senior2020heat, PhysRevApplied.15.054050}, thermal transistors \cite{PhysRevLett.111.063601,PhysRevLett.116.200601,PhysRevE.98.022118, liu2021common, PhysRevE.106.024110}, and thermometers \cite{PhysRevLett.114.220405, mukherjee2019enhanced} etc. Of particular interest for most functions
are thermal rectifiers, which play a critical role in the process of quantum information such as qubit initialization \cite{tuorila2017efficient}. The thermal diode as a two-terminal device mediated with temperature bias of two independent reservoirs can attain asymmetric heat fluxes. Such rectification effect has also been found in other systems \cite{PhysRevE.89.062109, PhysRevE.99.042121, PhysRevB.103.155434,motz2018rectification,PhysRevLett.120.200603, PhysRevE.99.032136, sanchez2015heat, PhysRevE.103.012134,PhysRevE.90.042142, tesser2022heat, senior2020heat, PhysRevE.95.022128}. Especially, in Ref. \cite{senior2020heat}, the author employs a qubit coupled to two superconducting resonators at different frequencies to realize the magnetic flux-tunable photonic heat rectification; In Refs. \cite{PhysRevE.89.062109, PhysRevE.95.022128},  the Ising interaction spins are employed to realize a well-performance based on the different excitation; In Ref. \cite{PhysRevE.104.054137}, the author proposed thermal rectification by utilizing the Dzyaloshinskii-Moriya interaction. It is obvious that the rectification effect in these models will vanish if one considers the resonant case and no other asymmetry is included. A usual understanding is that the rectification effect, or nonreciprocal heat transport, stems from some asymmetry of total system such as asymmetry of energy structure \cite{PhysRevE.89.062109} or nonlinear interaction \cite{motz2018rectification} or asymmetry couplings of system-reservoir \cite{ PhysRevApplied.15.054050, senior2020heat} and interactions \cite{PhysRevE.94.042135,PhysRevB.103.155434} or large anisotropy \cite{PhysRevLett.120.200603}  or asymmetry of the effective tunnelling barrier \cite{ghosh2022universal}. Whether some other features of the system induce (are closely related to) such nonreciprocal heat transport is still ambiguous now. In addition, despite the rapid advances in this field, it is still desirable that quantum thermal devices could be designed in terms of some experimentally friendly schemes.
\begin{figure}[!htbp] 
 \includegraphics[width=0.5\textwidth]{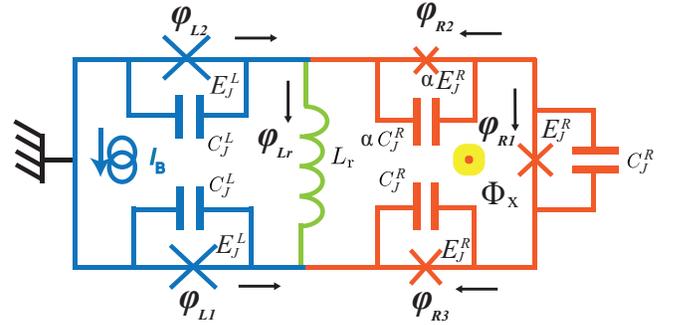} 
\caption{ The circuit QED architecture is modeled by the plasma mode of its direct-current superconducting quantum interference device (DC-SQUID) detector (left-hand side) inductively $L_{r}$ (middle) coupled to
a flux qubit \cite{PhysRevLett.101.070501} (right-hand side). DC-SQUID consists of the outer loop with the two large junctions \cite{PhysRevLett.63.1712, PhysRevLett.95.257002, PhysRevLett.79.2371} biased by current $I_{B}$ and two Josephson junctions have the same Josephson energy $E^{L}_{J}$ and capacitance $C^{L}_{J}$. In general, a flux qubit is the superconducting loop with
three-Josephson-junction \cite{mooij1999josephson}(shown with crosses) controlled by an external flux $\Phi_{\mathrm{x}}$, and two Josephson junctions have the same Josephson energy $E^{R}_{J}$ and capacitance $C^{R}_{J}$, and the other Josephson junction has a smaller Josephson energy $\alpha E^{R}_{J}$ as well as capacitance $\alpha C^{R}_{J}$ \cite{PhysRevB.60.15398}.}
\label{FIG1_} 
\end{figure}

Recently, circuit QED has a well-developed range from weak internal coupling to strong coupling even ultrastrong coupling\cite{RevModPhys.91.025005, frisk2019ultrastrong} with enormous design flexibility at microwave frequencies.
More importantly, circuit QED as an ideal experimental platform has been implemented to achieve the nonlinear ultrastrong coupling \cite{https://doi.org/10.48550/arxiv.cond-mat/0507290,PhysRevA.98.053859,PhysRevA.97.013851} between the qubit and resonator. Also, the nonlinear interaction has some interesting applications on the squeezing of light \cite{PhysRevA.47.3167}, spectral collapse \cite{PhysRevA.92.033817}, and multiphoton resonance \cite{PhysRevA.102.053709}. In addition, the generalized quantum Rabi model \cite{PhysRevA.102.053709} is also the fundamental form of cavity QED including one and two-photon interaction between atom and cavity which has been widely revisited. In this sense, the circuit QED system could be an appealing platform for experimentally friendly manipulation and is worthy of further exploration on thermal rectifiers. 
 
In this paper,  we will employ the circuit QED system to study the nonreciprocal heat transport and exploit quantum thermal diodes. We mainly study the two-photon Rabi model and reveal the steady-state nonreciprocal heat transport behaviors under different coupling conditions.  We first find out the perfect heat rectification in the quantum Rabi model and seek for good parameter regions for the experimental realization of thermal diodes. Most importantly, we comparably investigate the relation between heat rectification and photon detection rates in the dissipative compound system, which demonstrates similar nonreciprocal behaviors and indicates their potential relation. It provides a quantum optical perspective to understand quantum thermodynamic devices. The paper is organized as follows: In section \ref{section II}, the physical model and the master equation we employed are introduced. In section \ref{section III}, we study the effects of quantum thermal diodes and analyze the effects of different parameters. In section \ref{section IV}, we investigate the nonreciprocal photon detection rates. The conclusion and discussion are given in section \ref{section V}.

\section{Physical model and dynamics}
\label{section II}
We consider the two-photon quantum Rabi model including a  flux qubit coupling with a direct-current superconducting quantum interference device (DC-SQUID) via a small inductance with the circuit scheme sketched in Fig. \ref{FIG1_}. We would like to refer to the detailed derivation of the circuit model as Appendix A in Ref. \cite{PhysRevA.98.053859} and the references therein. The quantization Hamiltonian of the whole system reads 
 \cite{PhysRevA.98.053859,PhysRevA.97.013851, https://doi.org/10.48550/arxiv.cond-mat/0507290}(we assume $\hbar=1$) 
\begin{equation}
\tilde{H}_{S}=\omega_{L} a^{\dagger} a -\frac{1}{2}(q\sigma^{z}+\epsilon\sigma^{x})+g \sigma^{x}(a^{\dagger}+ a)^{2}, \label{equ:H_S1}
\end{equation}
where $a$, $a^{\dagger}$ are the creation and annihilation
operators of the SQUID, respectively, and $\sigma^{z}$, $\sigma^{x}$ are the Pauli matrices. The parameters $\omega_{L}$ denotes the frequency of the SQUID, $\epsilon$ and $q$ are the tunnel splitting and the energy separation of “persistent currents states”  \cite{PhysRevB.60.15398} induced by the external flux $\Phi_{\mathrm{x}}$ around the qubit loop as well as $g$ denotes the coupling strength between qubit and resonator. Here $q=2 I_{p} (\Phi_{\mathrm{x}}-\Phi_{\mathrm{0}}/2)$, $\Phi_{\mathrm{0}}=h/2e$ is the superconducting flux quantum and $I_{p}$ is the persistent current along the qubit loop \cite{mooij1999josephson}. 

The system inevitably dissipates due to the impedance $Z_{\omega}$ \cite{PhysRevB.81.144510, RevModPhys.93.025005}, which, serving as the environment, is described by an ensemble of harmonic oscillators \cite{PhysRevA.29.1419, PhysRevA.89.023817} with the free Hamiltonian
\begin{equation}
{H}_{R}=\sum_{\nu=L, R} \tilde{H}_{\nu}=\sum_{\nu l}\omega_{\nu l}{b}_{\nu l}^{\dagger}{b}_{\nu l}.\label{bath}
\end{equation}
In Eq. (\ref{bath}), $b_{\nu l}$, $b^{\dagger}_{\nu l}$ denote the annihilation and creation operators of the different reservoir mode with the  frequency $\omega_{\nu l}$.  Since the different types of noises originated from the fluctuations of distinct nature, the reservoirs we considered are independent and uncorrelated, i.e., $[b_{\nu l}, b^{\dagger}_{\mu j}]=\delta_{\nu \mu}\delta_{l j}$ \cite{PhysRevB.81.144510}. Thus the interaction between the system and the reservoirs can be described by a Caldeira-Leggett-type \cite{RevModPhys.59.1} Hamiltonian as 
\begin{equation} 
\tilde{H}_{SR}=\sum_{\nu l}\tilde{S}_{\nu} \otimes (\kappa_{\nu l}b_{\nu l}^{\dagger}+ \kappa^{*}_{\nu l}b_{\nu l}),\label{equ:H_SR}
\end{equation}
where 
\begin{align} \label{jump operator1}
\tilde{S}_{L}=a^{\dagger}+a, \ \tilde{S}_{R}=\sigma^{-}+\sigma^{+},
\end{align}
denote the dissipative transition operators of the system due to the coupling to the circuit environment \cite{PhysRevA.89.023817, PhysRevB.81.144510, shitara2021nonclassicality} and the parameter
$\kappa_{\nu l} $ reflects the interaction strength of the subsystem $\nu$ with its corresponding environment. For more physical insight \cite{PhysRevB.81.144510,PhysRevLett.121.060503},
one can rewrite the Hamiltonian Eq.~(\ref{equ:H_S1}) in the qubit eigenbasis by a rotation \begin{equation}
U=\left(\begin{array}{cc}
\rm \cos \theta/2 & -\rm sin \theta/2 \\
\rm sin \theta/2 & \rm \cos \theta/2
\end{array}\right)
\end{equation} as
 \begin{equation}
H_{S}=\omega_{L} a^{\dagger} a -\frac{1}{2}\omega_{R}\sigma_{z}+g (a^{\dagger}+ a)^{2} (\sigma_{z} \rm sin\theta+\sigma_{x} \rm cos\theta ), \label{equ:H_S}
\end{equation}
where $\tan \theta=\epsilon/q$, $\sigma_{z}=\rm\cos \rm\theta\sigma^{z}+\rm\sin \theta \sigma^{x}$ and, $\sigma_{x}=-\rm\sin \theta \sigma^{z}+\rm\cos \theta \sigma^{x}$. 
It is noted that $\omega_{R}=\sqrt{\epsilon^{2}+q^{2}}$ denotes the transition frequency of qubit.  
The corresponding system-reservoir interaction Hamiltonian Eq.~(\ref{equ:H_SR}) will be replaced by  \cite{PhysRevB.81.144510}
\begin{equation}
H_{SR}=\sum_{\nu l}\left(S_{\nu}\otimes (\kappa_{\nu l}b_{\nu l}^{\dagger}+ \kappa^{*}_{\nu l}b_{\nu l})\right),\label{equ:H_SR2}
\end{equation} with
\begin{align}
S_{L}=a^{\dagger}+a,\ S_{R}=\rm\sin \theta \sigma_{z}+\rm\cos \theta \sigma_{x}.\label{jump operator2}
\end{align}
Hence  the Hamiltonian of the total composite system including the reservoirs can be written as
\begin{equation}
H=H_{S}+H_{R}+H_{SR}.\label{equ:H_total}
\end{equation}

In order to get the dynamical equation of the system, we first write the Hamiltonian Eq.~(\ref{equ:H_S}) in the spectrum decomposition as \begin{equation}
H_{S}=\sum_{i}E_{i}\left|E_{i}\right\rangle \left\langle E_{i}\right|,
\end{equation}
where $E_{i}$ denotes the eigenvalue and $\left|E_{i}\right\rangle $
is its corresponding eigenstate. In the $H_{S}$ representation, the eigen-operators 
 are given by
\begin{equation}
\mathcal{S}_{\nu k}(\omega_{\nu k})=\sum_{E_{j}-E_{i}=\omega_{\nu k}}\left|E_{i}\right\rangle \left\langle E_{i}\left|S_{\nu}\right|E_{j}\right\rangle \left\langle E_{j}\right|,
\end{equation}
 which satisfies  the commutation relation 
\begin{equation}
\left[H_{S},~\mathcal{S}_{\nu k}(\omega_{\nu k})\right]=-{\nu k}\mathcal{S}_{\nu k}(\omega_{\nu k}).
\end{equation}
For convenience, we'd like to set $\mathcal{S}_{\nu k}=\mathcal{S}_{\nu k}(\omega_{\nu k})$. With these eigen-operators,  one can follow the standard process \cite{breuer2002theory, weiss2012quantum, PhysRevA.84.043832} to write   the master equation in the interaction picture  based on the Born-Markov-Secular approximation as
\begin{equation}
\frac{d \rho}{d t}=\sum_{\nu}\mathcal{L}_{\nu}[\rho]=\sum_{\nu k} \mathcal{L}_{\nu k}(\rho),\label{master equation}
\end{equation}
where the Lindblad dissipator $\mathcal{L}$ can be given as
\begin{align}
\nonumber \mathcal{L}_{\nu k}[x]=\Gamma^{\nu k}_{+}[\mathcal{S}_{\nu k}\rho \mathcal{S}_{\nu k}^{\dagger}-\frac{1}{2}\{\mathcal{S}_{\nu k}^{\dagger}\mathcal{S}_{\nu k},\rho\}]\\+\Gamma^{\nu k}_{-}(\omega_{\nu k})[\mathcal{S}_{\nu k}\rho \mathcal{S}_{\nu k}^{\dagger}-\frac{1}{2}\{\mathcal{S}_{\nu k}^{\dagger}\mathcal{S}_{\nu k},\rho\}],
\end{align}
where $\Gamma^{\nu k}_{+}=\Gamma^{\nu k}[\bar{n}(\omega_{\nu k})+1]$, $\Gamma^{\nu k}_{-}=\Gamma^{\nu k} \bar{n}(\omega_{\nu k})$ which denote the emission and absorption processes of an excitation, respectively \cite{PhysRevB.67.094510}, and $\bar{n}\left(w_{\nu k}\right)=\frac{1}{e^{\frac{\omega_{\nu k}}{T_{\nu}}}-1}$ denotes mean photon number of the mode  $\omega_{{\nu,k}}$ at temperature $T_{\nu}$ ($k_{B}=1$). In addition, with the Ohmic spectral densities, we have
the relaxation coefficients 
$\Gamma^{\nu k}=\gamma \omega_{\nu k}$  \cite{PhysRevLett.109.193602, PhysRevB.81.144510}. 

\section{Quantum thermal diode}  \label{section III}
We first transform the Eq. (\ref{master equation}) into Schr\"{o}dinger
picture,
\begin{equation}
\frac{d \rho}{d t}=-i[H_S, \rho]+\sum_{\nu}\mathcal{L}_{\nu}[\rho].\label{master equation in the S-picture}
\end{equation}
A direct expansion of Eq. (\ref{master equation in the S-picture}) can show that the diagonal and off diagonal entries of the density matrix are decoupled to each other and especially the off diagonal entries will vanish for steady state. Hence we only address the evolution of the diagonal entries of $\rho$ as \cite{PhysRevE.90.052142, PhysRevE.89.062109}, i.e., 
\begin{equation} \label{the evolution of diagonal elements}
\dot{\rho}_{k k}=\sum_{\nu=L, R} \sum_l \Gamma_{k l}^\nu(\rho),
\end{equation}
where $\Gamma^{\nu}_{i j}(\rho)=\left(\Gamma^{\nu k}_+ \rho_{j j}-\Gamma^{\nu k}_- \rho_{i i}\right)\left|\left\langle E_i\left|S_v\right| E_j\right\rangle\right|^2$ denote the net decay rate from the state $|E_j\rangle$ to state $|E_i\rangle$ induced by interacting with the $\nu$ reservoir. The diagonal steady-state density matrix $\rho_{s}$ of Eq. (\ref{the evolution of diagonal elements}) can be obtained by solving $\dot{\rho}_{k k}=0$, i.e.,
\begin{equation}
\sum_{\nu=L, R} \sum_l \Gamma_{k l}^\nu(\rho^s)=\mathcal{M}\left\vert\rho_{s}\right\rangle=0,\label{juzhen}
\end{equation} 
where $\left\vert\rho_{s}\right\rangle$ denotes the columns vector consisting of the diagonal entries of $\rho^s$ and $\mathcal{M}$ is the corresponding coefficient matrix obtained from Eq. (\ref{juzhen}). Thus the steady-state $\rho_{s}$ can be solved by the null space of the linear operator $\mathcal{M}$.

Based on the steady state $\rho_{s}$, one can directly calculate the steady-state heat current which is defined by \cite{breuer2002theory},
\begin{equation} \label{definition of heat current}
\dot{\mathcal{Q}}_{\nu}=\operatorname{Tr}\left\lbrace H_{S}\mathcal{L}_{\nu}\left[\rho^s\right]\right\rbrace. 
\end{equation} 
$\dot{\mathcal{Q}}_{\nu}>0$ denotes heat flows from the reservoir to the system and $\dot{\mathcal{Q}}_{\nu}<0$ denotes heat flows from the system into the reservoir.  
With Eq. (\ref{the evolution of diagonal elements}), the heat current Eq. (\ref{definition of heat current}) can be explicitly written as
\begin{equation} \label{definition of heat current at SS}
\dot{\mathcal{Q}}_{\nu}=-\sum_{k l}
\Gamma^{\nu}_{k l} (\rho_{s}) E^{\nu}_{k l},
\end{equation} 
where $E^{\nu}_{k l}=E_{l}-E_{k}$ denotes the transition frequency. Substituting the heat currents (\ref{definition of heat current}) into Eq. (\ref{juzhen}), one can easily check that the two heat currents fulfill the conservation relation  $\dot{\mathcal{Q}}_{L}+\dot{\mathcal{Q}}_{R}=0$.  Due to the large dimension of our system, the analytical computation is quite hard and we can only approximately solve the question with truncating  photon numbers $N=2$ in the low temperature regime as in Appendix \ref{Appendix B}. Therefore, to reveal the behaviors of heat transport in the system, we will numerically \cite{JOHANSSON20131234} solve the steady state of the master equation (\ref{master equation in the S-picture}) as well as the steady-state heat currents. 
 At first, we mainly take two coupling regimes as examples, one is the strong coupling with  $g=0.015\omega_0$, and the other is the ultrastrong coupling with $g=0.45\omega_0$. In the numerical process, the dissipation rates are taken as $\gamma=10^{-4} \omega_{0}$ and the resonator frequency takes $\omega_{L}=\omega_{0}$. The relevant parameters are given in Table.~\ref{table1} \cite{mooij1999josephson,PhysRevB.76.174523,yan2016flux,tan2017quantum,PhysRevLett.118.103602,yoshihara2017superconducting} if not specified.  
 \begin{figure}[!htbp]
\centering \includegraphics[width=1.0\columnwidth]{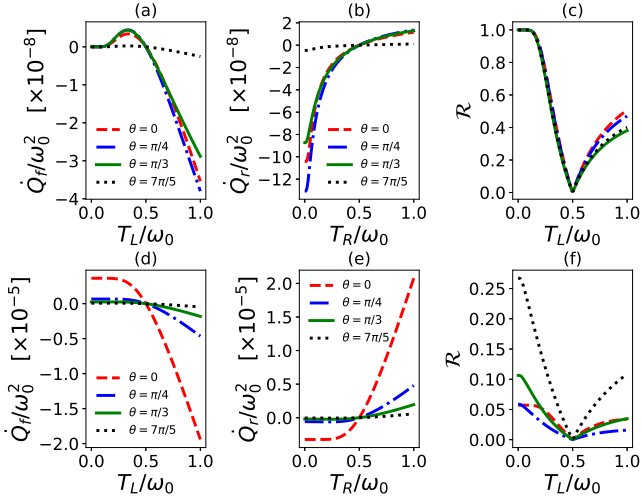} \caption{Heat currents (a-b, d-e) and rectification coefficients (c, f) versus the temperatures of reservoirs for qubit-resonator coupling strength $g=0.015\omega_{0}$.  The top panel denotes the non-resonant case with qubit frequency $\omega_{R}=0.1\omega_{0}$ and the bottom panel denotes the two-photon resonant case with qubit frequency $\omega_{R}=2\omega_{0}$. The dissipation rates are considered as $\gamma=10^{-4}\omega_{0}$ and resonator frequency $\omega_{L}=\omega_{0}$. It is noted that in the forward (reverse) heat transport process, we fix the temperature $T_{R} (T_{L})=0.5 \omega_{0}$ and the related realistic parameters are shown in Table \ref{table1}. 
}
\label{FIGURE_hc_sc} 
\end{figure}
 
\begin{table}[th]
\begin{tabular}{|ccc|}\hline
  & Simulation parameters  &  \\ \hline
Symbol & Parameter  & Value\\ \hline
$\omega_{\rm 0}/2\pi$ & reference frequency & $20 \mathrm{GHZ}$\\
$\omega_{\rm L}/2\pi$ & cavity frequency & $20 \mathrm{GHZ}$\\
$\omega_{\rm R}/2\pi$ & qubit frequency & $- \mathrm{GHZ}$\\
$g/2\pi$ & coupling strength & $- \mathrm{GHZ}$\\
$\gamma/ 2 \pi$ & dissipation rate & $2 \mathrm{MHZ}$\\
$T_{\rm R}$ ($T_{\rm L}$) & temperature & $960\mathrm{mK}$\\
\hline
\end{tabular}
\caption{Parameters of two-photon quantum Rabi system. }
\label{table1}
\end{table} 
\begin{figure}[!htbp]
\centering \includegraphics[width=1.0\columnwidth]{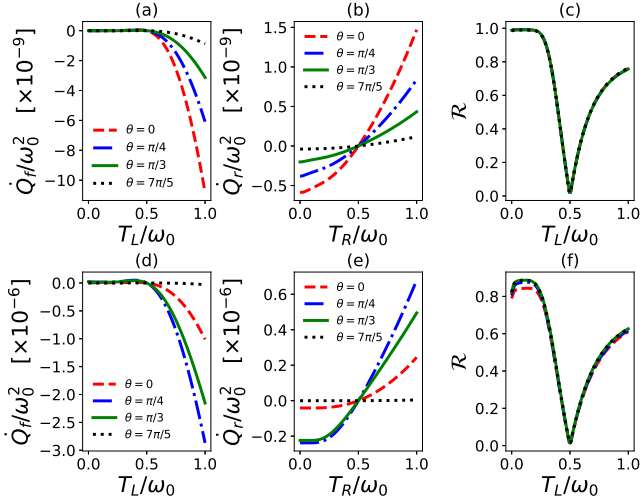} \caption{Heat currents (a-b, d-e) and rectification coefficients (c, f) versus the temperatures of reservoirs for qubit-resonator interaction strength $g=0.45\omega_{0}$. The top panel corresponds to off-resonance case for qubit frequency $\omega_{R}=0.1\omega_{0}$ and the bottom panel correspond to two-photon resonance case for qubit frequency $\omega_{R}=2\omega_{0}$. Other parameters are the same as those in Fig. \ref{FIGURE_hc_sc}.
}
\label{FIGURE_hc_usc} 
\end{figure}

For clarity, we define $\dot{\mathcal{Q}}_{f/r}$ as the forward (reverse) heat currents since we need to exchange the temperatures of two terminals for the demonstration of nonreciprocal heat transports. $\dot{\mathcal{Q}}_{f}=\dot{\mathcal{Q}}_{R}$ denotes heat current from the $R$ reservoir to the system when $T_{R}>T_{L}$, and vice verse. The intuitive illustrations of the heat currents through the system for the two coupling regimes are given  in Figs. \ref{FIGURE_hc_sc} (a-b, d-e) and Figs. \ref{FIGURE_hc_usc} (a-b, d-e), respectively, where we keep $T_R=0.5\omega_0$ in subfigures (a), (d) and  $T_L=0.5\omega_0$ in subfigures (b), (e).  One can find from the figures that the heat current vanishes when the two terminals have the same temperature $0.5\omega_0$, which coincides with our intuitive understanding. In particular, these figures indicate the apparent nonreciprocal heat transport.  Namely, with one terminal temperature fixed, the relatively large heat current can only appear in a single direction. The subfigures (a), (d) compared with the corresponding subfigures (b), and (e) imply the exchange of the temperatures of two terminals. Therefore, by comparing the pair of subfigures (a) and (b), or (d) and (e) in both Fig. \ref{FIGURE_hc_sc} and Fig. \ref{FIGURE_hc_usc},  one can find the nonreciprocal heat transport more clearly in both resonant and non-resonant cases. This means that the current system can be designed as a thermal diode.

A key index to characterize the performance of a thermal diode is the rectification coefficient which is defined by \cite{PhysRevE.90.042142}  
\begin{equation}
\mathcal{R}=\frac{\left|\dot{\mathcal{Q}}_{f}+\dot{\mathcal{Q}}_{r}\right|}{|\dot{\mathcal{Q}}_{f}-\dot{\mathcal{Q}}_{r}| },
\end{equation}
where $\mathcal{R}$=1 for perfect diode, $0<\mathcal{R}<1$ for a good diode, however, $\mathcal{R}$=0 for no rectification when $\dot{\mathcal{Q}}_{r}=-\dot{\mathcal{Q}}_{f}$. In Figs. \ref{FIGURE_hc_sc} and \ref{FIGURE_hc_usc} (c) and (f), we plot the rectification coefficient depending on the temperature. The rectification effect is gradually enhanced with the deviation from the equilibrium temperature $T_R=T_L=0.5\omega_0$. All the figures demonstrate certain rectification effects, but the two-photon resonant weak coupling allows quite weak rectification, and good performance of the diode can be found for the non-resonant strong coupling ($\omega_R=0.1\omega_0$, $g=0.015\omega_0$) and the (resonant and nonresonant) ultrastrong coupling $g=0.45\omega_0$).  In particular, $\theta$ can obviously affect the heat currents, but it only slightly affects the rectification coefficients. In the sense of a heat diode, one can adjust heat currents by $\theta$ according to the practical requirement simultaneously without greatly disturbing the rectification performance. 

\begin{figure}[!htbp]
\centering \includegraphics[width=1.0\columnwidth]{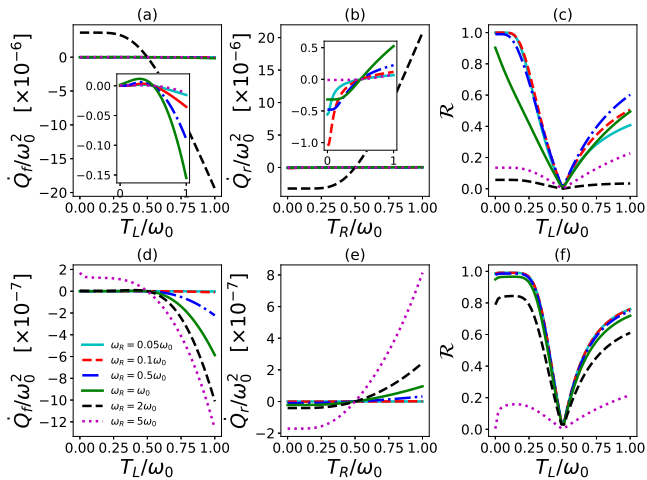} \caption{ Heat currents (a-b, d-e) and rectification coefficients (c, f) as a function of the temperature $T_{L/R}$ for different qubit frequencies.  In the top panel, qubit-resonator coupling $g=0.015\omega_{0}$; In the bottom panel,  $g=0.45\omega_{0}$. The inset views (a-b) show that heat currents as a function of temperature $T_{L/R}$ other than two-photon resonant case with qubit frequency $\omega_{R}=2 \omega_{0}$. Here $\theta=0$ and other parameters are the same as those in Fig. \ref{FIGURE_hc_sc}. } 
\label{FIG_hc_TL_omegaL} 
\end{figure}

As analyzed above, the rectification coefficients are relatively small in the case of resonant strong coupling $g=0.015\omega_0$. A rough conclusion is that the ultrastrong coupling or the non-resonant coupling could be beneficial to the performance of a heat diode.  However, a detailed investigation reveals that for $g=0.015\omega_0$  in Figs. \ref{FIG_hc_TL_omegaL} (a-c), the heat currents are relatively large around the resonant coupling and gradually decrease with deviating from resonant coupling, but the rectification coefficients take the relatively large values around $\omega_R=0.1\omega_0$. For $g=0.45\omega_0$, we find that the heat currents increase from $\omega_R=0.05\omega_0$ to  $\omega_R=5\omega_0$ in Figs. \ref{FIG_hc_TL_omegaL} (d-f), but the best rectification coefficients appear around $\omega_R=0.05\omega_0$. 

\begin{figure}[!htbp]
\centering \includegraphics[width=1.0\columnwidth]{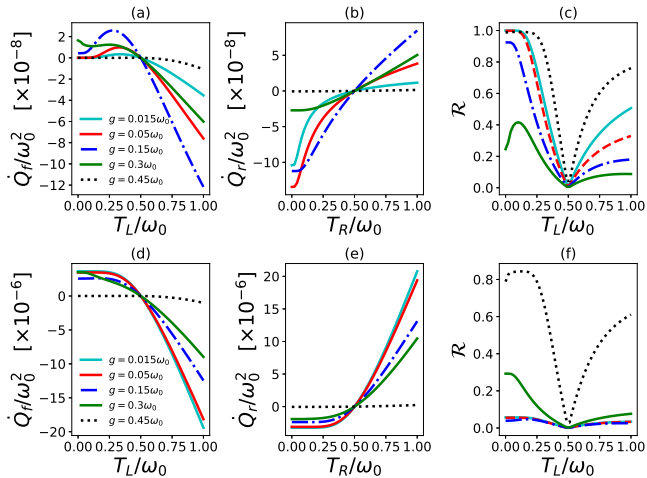} \caption{Heat currents (a-b, d-e) and rectification coefficients (c-f) versus the temperautes for different qubit-resonator coupling strengths. In the top panel, considering the non-resonant condition with qubit frequency $\omega_{R}=0.1\omega_{0}$; In the bottom panel, considering two-photon photon resonant condition with $\omega_{R}=2\omega_{0}$.  Here $\theta=0$ and other parameters are taken as Fig. \ref{FIGURE_hc_sc}.}
\label{FIG_hc_TL_g} 
\end{figure}

In addition, we also study the effect of the coupling strength $g=0.015\omega_0,0.05\omega_0,0.15\omega_0,0.3\omega_0,0.45\omega_0$ on the performance of the heat diode with fixed $\omega_L=\omega_0$ shown in Fig. \ref{FIG_hc_TL_g}. The results indicate that not only the heat currents but also the rectification coefficients don't monotonically depend on the coupling, but the best rectification coefficients are achieved by the ultrastrong coupling $g=0.45\omega_0$.
In one word, for a given $\omega_L=\omega_0$, the appropriately large red detuning for $\omega_R$ seems to be much better for a good heat diode in both coupling regimes, and the ultrastrong coupling is more conducive to the heat diode. Comparing Fig. \ref{FIG_hc_TL_g} (a-b) and (d-e), we find that heat current at $g=0.45 \omega_{0}$ both resonant and nonresonant cases are relatively smaller and this system can behave a well-performance diode.
Besides the interaction strength, the coupling mechanism plays an important role. To give a comparison, we also study the rectification effect of different coupling mechanisms with various coupling strengths in Appendix \ref{Appendix A}. It can be found that different coupling mechanisms can achieve certain rectification of heat currents, but the current Rabi model apparently has a better rectification effect in the ultrastrong coupling regime than in other cases.

An intuitive understanding of the nonreciprocal heat transport can be attributed to the asymmetric transition rates induced by the asymmetric structure of the system. Of course, it could be hard for a system with many energy levels to give a brief and precise picture, but a rough picture can always be found from the asymmetric coupling with the environments. In Appendix \ref{Appendix B}, we have drawn the allowed transition diagram and plotted the corresponding transition rates. From the diagram, one can easily see that the allowed transitions coupled to different environments are quite different, which usually produces large variations of transition rates if exchanging two different temperatures. This can be further verified by the figures for transition rates. The transition rates shown in the figures are distributed along both sides of the $x$ axis, the summation of which forms the total heat current.  Comparing Figs. \ref{FIGURE_transition_rates} (a) and (b), it is clear that exchanging the temperatures will greatly change the magnitude distributions of transition rates, which further determines the different heat currents, i.e., the nonreciprocal heat transport. As a comparison, in the resonant coupling case given in Figs. \ref{FIGURE_transition_rates} (c) and (d), exchanging the temperature leads to a little bit similar magnitude distributions, which means weak nonreciprocity. As for the ultrastrong coupling regime $g=0.45 \omega_{0}$, we only consider the transitions of the four lowest states of the model as shown in Fig. \ref{transition rates in ultrastrong regime}. From this figure, we find the transition cycles for resonant and non-resonant cases can produce nonreciprocal heat transport.  Such a picture should also be applicable in other schemes of thermal diodes.

\section{Nonreciprocal photon detection rates} 
\label{section IV}

The thermal diode is one result of the nonreciprocal behavior of thermal transfer, which can be usually understood from some asymmetric features of the system. On the contrary, such an asymmetric feature could induce versatile nonreciprocal behaviors of different perspectives, thus it forms a bridge between different physical phenomena.  In other words, one could understand the thermal diode from a different angle. Physically, the quantum thermal diode works by absorbing net numbers of photons from one side and release to the side, so the first intuition is that the thermal diode could be related to the nonreciprocity of the output photon numbers. As we have analysed, the asymmetric structure of SQUID-qubit-reservoir leads to the unidirectional heat transport and it also reflect the asymmetry of photon detection rates. As an important result, we'd like to study the potential link between thermal and optical behaviors.

\begin{figure}[!htbp]
\centering \includegraphics[width=1.0\columnwidth]{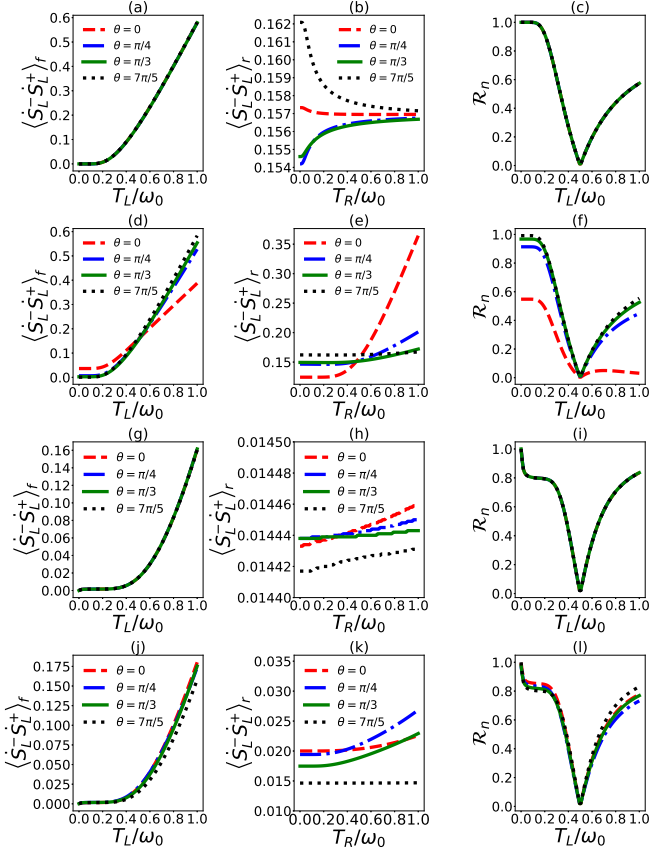} \caption{Photon fluxes and asymmetry coefficients versus the temperatures with different angles $\theta$ for non-resonant cases (a-c; g-i) and two-photon resonant cases (d-f; i-l). (a)-(f) correspond to $g=0.015\omega_{0}$, and (g)-(l) correspond to $g=0.45\omega_{0}$. Other parameters are the same as Fig. \ref{FIGURE_hc_sc}. 
}
\label{FIG_photon_flux_theta} 
\end{figure}
As we know,  in the strong coupling regime  \cite{PhysRevA.80.053810} the usual input-output relations have to be generalized. Here we directly start with
the input-output relation \cite{PhysRevLett.109.193602}  \begin{equation} \label{input-output}
b_{\text {out }}(t)=b_{\text {in }}(t)+ \sqrt{\gamma} \dot{S}_{\nu}^{+},\end{equation}
where $\sqrt{\gamma}=\frac{\epsilon_{c}}{\sqrt{4 \pi \epsilon_{o} v}}$, $\dot{S}_{\nu}^{+}=-i \sum_{i, j>i} \omega_{\nu k} \mathcal{S}_{\nu k}$. $b_{in}$ denotes the input field operator,
\begin{equation}
b_{in} =\frac{1}{\sqrt{2 \pi}}\sum_{l}  \sqrt{\omega_{\nu l}} e^{-i \omega_{\nu l} (t-t_0)} b_{\nu l} (t_0),
\end{equation}
and $b_{out} (t)$ correspond to the output field  operator,
\begin{equation}
b_{out} (t)=\frac{1}{\sqrt{2 \pi}}\sum_{l}  \sqrt{\omega_{\nu l}} e^{-i \omega_{\nu l} (t-t_1)} b_{\nu l} (t_1).
\end{equation}
In addition, in the setting of Ref. \cite{PhysRevLett.109.193602}$, \epsilon_c$ is a coupling parameter of the cavity field and the environment (waveguide field outside the cavity) and $\epsilon_o$ is a parameter describing the dielectric
properties of the output waveguide, and $v$ is the phase velocity. Thus one can measure the output ac-voltage in circuit QED which is proportional to the mean output photon number $\left\langle b^\dagger_{\text {out }}(t)b_{\text {out }}(t)\right\rangle$. 
In particular, if there is no input field,  the output field is proportional to $\dot{S}_{\nu}^{+}$, namely,  $\left\langle\dot{S}_{L}^{-} \dot{S}_{L}^{+}\right\rangle$ can be used to study the mean photon number.

\begin{figure}[!htbp]
\centering \includegraphics[width=1.0\columnwidth]{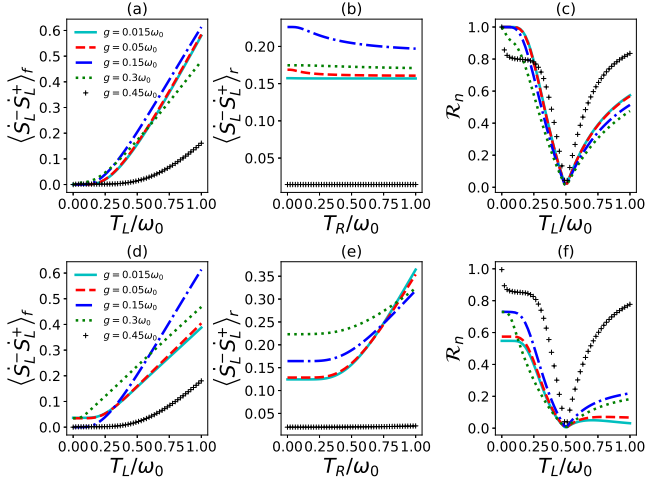} \caption{Photon fluxes (a-b, d-e) and asymmetry coefficients (c, f) for different coupling strengths. We consider the qubit frequency with $\omega_{R}=0.1\omega_{0}$ in the upper panels and $\omega_{R}=2\omega_{0}$ in the lower panels. The angle $\theta=0$ and other parameters are taken the same as Fig. \ref{FIGURE_hc_sc}.}
\label{photon_flux_g} 
\end{figure}

\begin{figure}[!htbp]
\centering \includegraphics[width=1.0\columnwidth]{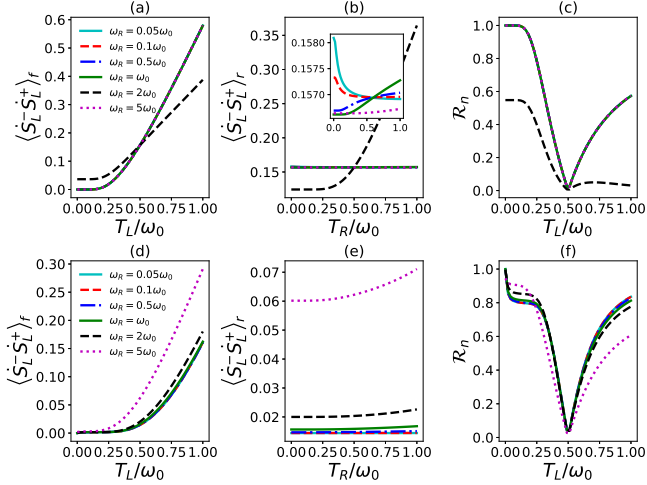} \caption{Photon fluxes (a-b, d-e) and asymmetry coefficients (c-f) as a function of temperatures for different qubit frequencies. The inset view (b) shows that heat currents as a function of temperature $T_{R}$ other than two-photon resonant case $\omega_{R}=2 \omega_{0}$. We set the qubit-resonator interaction strength $g=0.45\omega_{0}$, and $\theta=0$. Other parameters are taken the same as Fig. \ref{FIGURE_hc_sc}.
}
\label{FIG_photon_TL_omegaR} 
\end{figure}

Similarly, to describe the asymmetry of the mean photon number subject to the forward and reverse directions, we  define
the asymmetry coefficient  in the long time limit as
\begin{equation}
\mathcal{R}_{n}=\frac{\left| \left\langle\dot{S}_{L}^{-} \dot{S}_{L}^{+}\right\rangle_f-\left\langle\dot{S}_{L}^{-} \dot{S}_{L}^{+}\right\rangle_{r}\right|}{  |\left\langle\dot{S}_{L}^{-} \dot{S}_{L}^{+}\right\rangle_f+\left\langle\dot{S}_{L}^{-} \dot{S}_{L}^{+}\right\rangle_r| }
\end{equation}
with the subscript $f$ denoting the forward direction and $r$ representing the reverse direction. When the coefficient $\mathcal{R}_{n}=0$ indicates the symmetry and the  $\mathcal{R}_{n}=1$ shows the maximal asymmetry. In Figs. \ref{FIG_photon_flux_theta}, \ref{photon_flux_g}, and \ref{FIG_photon_TL_omegaR}, we plot the mean photon numbers under different conditions and the corresponding asymmetry coefficients. All the figures indicate the nonreciprocal behaviors of mean photon numbers if the two terminals are exchanged. The mean photon numbers exhibit symmetric behaviors at the equilibrium temperature $T_L=0.5\omega_0$, while the strong asymmetry emerges far away from the equilibrium temperature, especially in the lower temperature region of $T_L$. In particular, in the strong coupling regime, the behaviors of asymmetric coefficients with the temperature $T_L$  are quite similar to the asymmetric behaviors of the heat currents. Although the ultrastrong coupling regime demonstrates a slight difference at the lower temperature region from the heat currents, the rough trends, i.e., asymmetry increasing with the decrease of $T_L<0.5\omega_0$, are completely consistent with each other. It is also interesting that similar to the heat currents, $\theta$ has a great influence on the mean photon numbers, but the influence on the asymmetric coefficients is not so apparent. Besides, we also illustrate the nonreciprocal mean photon numbers with different coupling strengths and frequencies taken into account, all show similar nonreciprocal behaviors to heat currents. In this sense, the asymmetry of the mean photon numbers in the current model could shed new light on the thermal diode behaviors.

\section{Conclusions and discussion}
\label{section V}
Before the end, we'd like to emphasize that the thermal diode is designed completely based on the experimentally friendly circuit QED platform and all the numerical process are considered with the  realistic parameters as shown in Table \ref{table1} and those in Ref. \cite{PhysRevA.98.053859}, i.e., $\alpha=0.8$, $E^{L}_{C}=2\times 10^{-3} E^{L}_{J}$, $L_{r}=30 E^{L}_{J}$, $E^{R}_{J}=11.6 E^{L}_{J}$, $E^{R}_{C}=E^{R}_{J}/80$.

In conclusion, we consider a two-photon quantum Rabi model as a perfect rectifier and discuss its corresponding properties of photon detection rates. It is shown that the ultrastrong coupling makes this model become a well-performance thermal diode. In addition, the non-resonant coupling within a certain range also plays an active role. Usually, the quantum thermal diode phenomenon can be intuitively understood in terms of the asymmetric structure of the system, which further induces asymmetric transition rates. We also study the nonreciprocity of the steady-state output photon detection rates, which exhibits quite similar nonreciprocal behavior to heat currents. Therefore, this paper could provide a new perspective to understand quantum thermal diodes. Simultaneously, it could also pave the way for future research on the thermodynamical and optical phenomena of multiqubit two‑photon Rabi model.

\section*{Acknowledgement}
This work was supported by the National Natural Science Foundation of China under Grant  No.12175029, No. 12011530014 and No.11775040, and the Key Research and Development Project of Liaoning Province, under grant 2020JH2/10500003.

\appendix
\section{Comparison of different coupling mechanisms} \label{Appendix A}
\begin{figure}[!htbp]
\includegraphics[width=0.95\columnwidth]{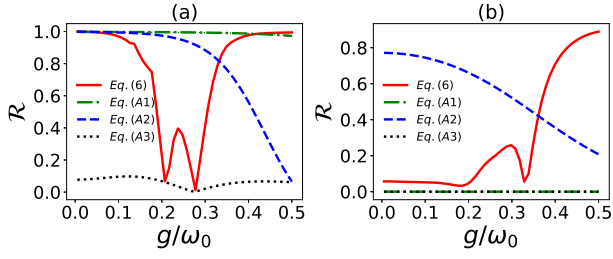} 
 \caption{Rectification coefficients for different coupling strengths. In the left panel, $\omega_{R}=0.1\omega_{0}$ for off-resonant case; in the right panel, for two-photon resonant case $\omega_{R}=2\omega_{0}$ with Eq. (\ref{equ:H_S}), while for resonant case $\omega_{R}=\omega_{0}$ with Eqs. (\ref{eq. Ising-interaction}), (\ref{eq. asymmery-interaction}), and (\ref{eq. DM-interaction}). Other parameters are taken as $\omega_{L}=\omega_{0}$, and $\theta=0$ as well as $T_{L}=0.1 \omega_{0}$ $T_{R}=0.5 \omega_{0}$ for forward process and vice verse. 
}
\label{different_models} 
\end{figure}

Considering the heat rectification effects, many different types of coupling mechanisms have been proposed. For example, in Refs. \cite{PhysRevE.89.062109, PhysRevE.95.022128}, the Hamiltonian reads
\begin{equation} \label{eq. Ising-interaction}
H_{S}=\frac{1}{2} (\omega_{L} \sigma^{L}_{z}+ \omega_{R} \sigma^{R}_{z} + g \sigma^{L}_{z} \sigma^{R}_{z}),
\end{equation}
and  Ref. \cite{PhysRevE.99.042121} takes the interaction as the following form
\begin{equation} \label{eq. asymmery-interaction}
H_{S}=\frac{1}{2} \omega_{L} \sigma^{L}_{z}+ \frac{1}{2} \omega_{R} \sigma^{R}_{z} +g \sigma^{L}_{z} \sigma^{R}_{x}.
\end{equation}
In addition, two qubits are coupled with Dzyaloshinskii-Moriya interaction as \cite{PhysRevE.104.054137} 
\begin{equation} \label{eq. DM-interaction}
H_{S}=\frac{1}{2} \omega_{L} \sigma^{L}_{z}+ \frac{1}{2} \omega_{R} \sigma^{R}_{z} +g (\sigma^{L}_{x} \sigma^{R}_{y}-\sigma^{L}_{y} \sigma^{R}_{x}).
\end{equation}
The form of system-reservoir coupling about the above Hamiltonian can be captured by
\begin{equation}
H_{S-R}=\sum_{\nu l}\left(X_{\nu}\otimes (\kappa_{\nu l}b_{\nu l}^{\dagger}+ \kappa^{*}_{\nu l}b_{\nu l})\right),
\end{equation}
where $X_{\nu}=\sigma^{+}_{\nu}+\sigma^{-}_{\nu}$ are the system operators, and $\sigma^{\pm}_{\nu}$ denote the spin operators.
We employ the global master equation and reproduce the results of Ref. \cite{PhysRevE.89.062109, PhysRevE.95.022128, PhysRevE.99.042121, PhysRevE.104.054137} for resonant and non-resonant cases with the same parameter conditions.
In Fig. \ref{different_models}, we illustrate the rectification coefficients versus coupling strength with different coupling mechanisms taken into account. It can be found that different 
coupling mechanisms have their own advantages under their particular parameter conditions. The two-photon Rabi mechanism can achieve good heat rectification effects in the ultrastrong coupling regime and exhibit certain robustness on the frequency matching.
 \section{Transition rates} \label{Appendix B}
\begin{figure}[!htbp]
\centering \includegraphics[width=0.85\columnwidth]{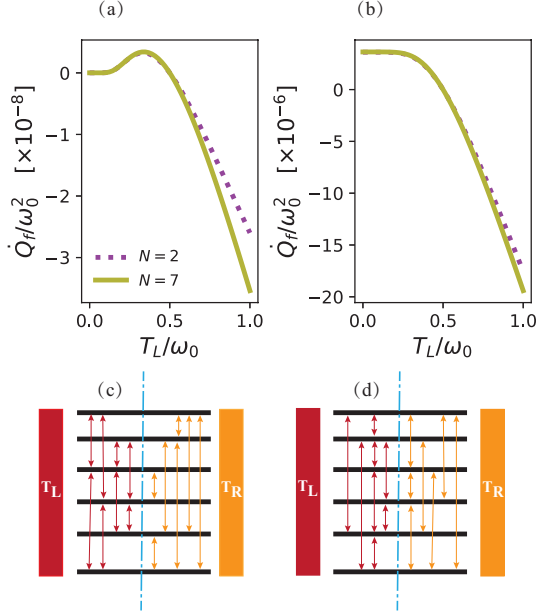} \caption{ Heat currents for the different truncations of photon numbers. It is shown that the heat current for the cavity truncation $N=2$ is approximately accurate in the low temperature considering the non-resonant (a) and two-photon resonant (b) cases. Here we take $g=0.015\omega_{0}$. The allowed transitions of the system subject to the left and right baths are shown in (c) and  (d).  The left panels are the non-resonant case with $\omega_{R}=0.1\omega_{0}$ and the right panels are the two-photon resonant case with $\omega_{R}=2\omega_{0}$. Other parameters are the same as those in Fig. \ref{FIGURE_hc_sc}. }
\label{FIGURE_hc_trun} 
\end{figure}

In order to give a physical understanding, one will have to analyze the energy level structure and the transition rates. However, as the complexity of the system structure increases, an integral and precise picture seems hard to be obtained. So we restrict the cavity truncation to two photons to present a simple picture. The diagram is given in Fig. \ref{FIGURE_hc_trun}, which indicates the validity of our simplicity as well as the allowed transitions subject to different environments. One can find that allowed transitions subject to the left and the right baths are quite different, which is the root of the nonreciprocal heat transport. They directly lead to asymmetric transition rates, which will be analyzed later. 

The heat current for the left terminal in the non-resonant case can be written in terms of Eq. (\ref{definition of heat current at SS}) as
\begin{eqnarray}
\dot{\mathcal{Q}}_{L}&=-\sum\limits_{i=1}^2\sum\limits_{j=3}^4\left(\Gamma^{L}_{ij} E_{ij}+\Gamma^{L}_{i+2,j+2} E_{i+2,j+2}\right),
\end{eqnarray}
where $E_{ij}=E_{j}-E_{i}$, and  the forward heat current reads $\dot{\mathcal{Q}}_{f}=-\dot{\mathcal{Q}}_{L}$ defined as before.
\begin{figure}[!htbp] \label{transition rates}
\centering \includegraphics[width=0.95\columnwidth]
{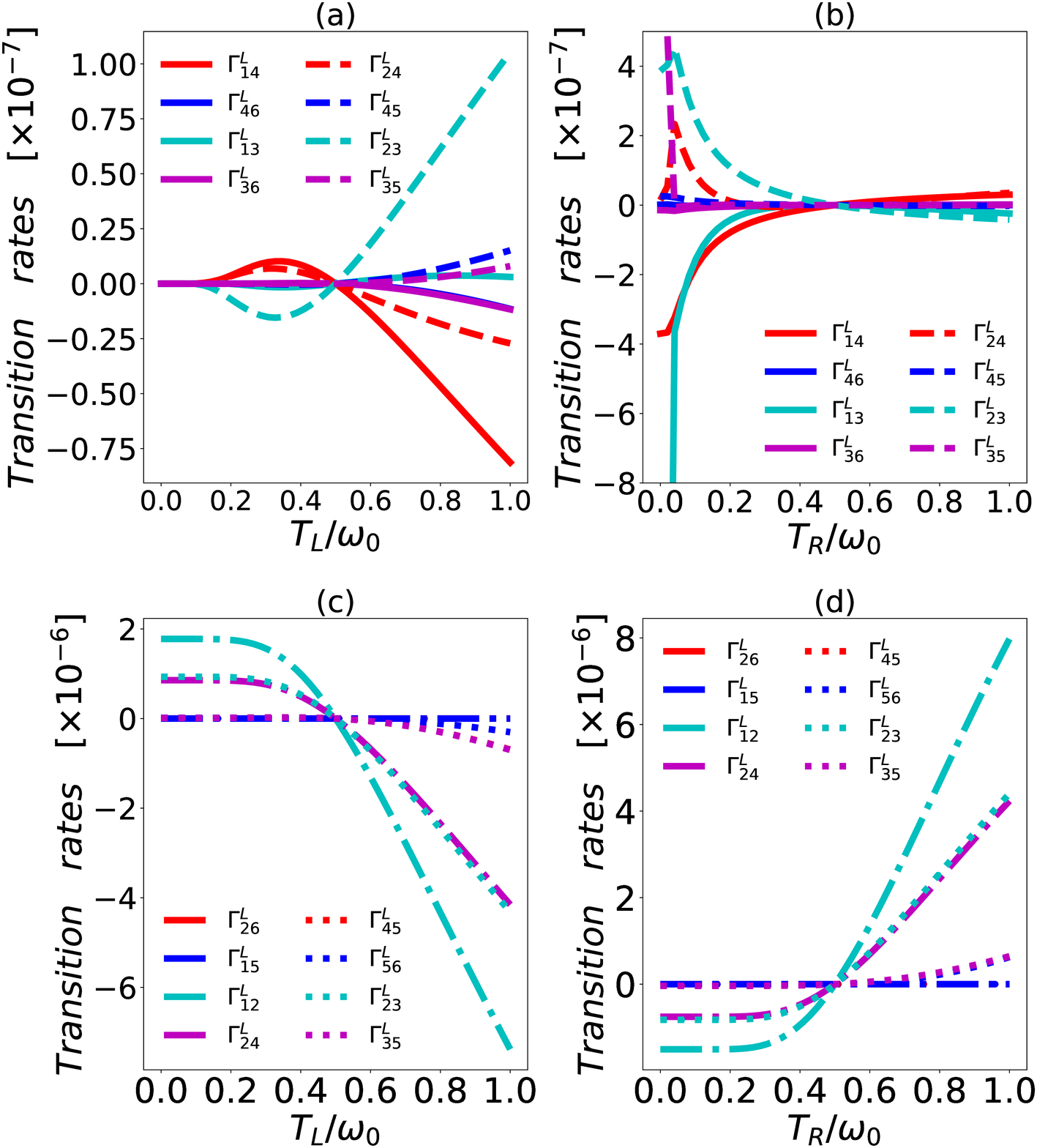}
\caption{The transition rates of the forward and reverse processes versus the temperatures for non-resonant case (a), (b) ($\omega_{R}=0.1\omega_{0}$) and resonant case (c), (d) ($\omega_{R}=2\omega_{0}$). It is noted that the cavity mode is truncated with $N=2$. The left panel corresponds to the forward process and the right panel to the reverse process. Here $g=0.015\omega_{0}$ and other parameters are the same as Fig. \ref{FIGURE_hc_sc}
}
\label{FIGURE_transition_rates} 
\end{figure}

Similarly, the heat current in the resonant case can be expressed as 
\begin{eqnarray}
\dot{\mathcal{Q}}_{L}&=-\sum\limits_{\substack{i=1\\i\neq 2,5}}^6\left(\Gamma^{L}_{i2} E_{i2}+\Gamma^{L}_{i5} E_{i5}\right).
\end{eqnarray}
According to the above expressions, we can plot the corresponding transition rates as a function of the temperature $T_{L}$ as shown in Fig. \ref{FIGURE_transition_rates}.  From the figures, one can see that the transition rates corresponding to the different allowed transitions in Fig. \ref{FIGURE_hc_trun} are distributed at the upper and lower sides along the $x$ axis. The contributions of the two sides are opposite, and all the transition rates associated with the corresponding transition energies together contribute to the given heat current. Comparing Figs. \ref{FIGURE_transition_rates} (a) and (b), one can see a typical characteristic that the magnitude distribution of the transition rates is quite different. This can be regarded as the most direct reason for the nonreciprocal heat transport. However, comparing Figs. \ref{FIGURE_transition_rates} (c) and (d), one can find that the transition rates are distributed a little bit similarly (neglecting the positive or negative rates), which implies relatively weak nonreciprocity. This has been clearly illustrated by Fig. \ref{FIGURE_hc_sc}.

\begin{figure}[!htbp] 
\centering \includegraphics[width=0.85\columnwidth]{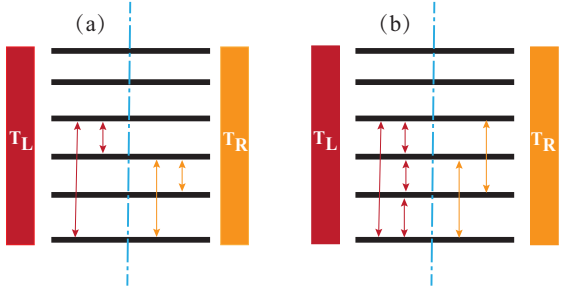} \caption{ The energy eigenvalues of the four lowest states of the two-photon Rabi model for non-resonant (a) and resonant (b) cases. Here we take $g=0.45\omega_{0}$ and $\omega_{L}=\omega_{0}$. The allowed transitions of the system subject to the left and right baths are shown in different color arrows. The qubit frequencies  are $\omega_{R}=0.1\omega_{0}$ for non-resonant case and $\omega_{R}=2\omega_{0}$ for the two-photon resonant case. }
\label{transition rates in ultrastrong regime}
\end{figure}

\section{The input-output formalism} 
From Eq. (\ref{equ:H_total}), we follow the method of  Ref. \cite{PhysRevLett.109.193602, le2020theoretical, girvin2011circuit} and derive the input-output relation.
The total Hamiltonian of the system and reservoir including their interaction is captured as 
\begin{align}
H=H_{S}+H_{R}+H_{SR},
\end{align}
\begin{align}
H_{SR}=&\sum_{\nu l}\left(S_{\nu}\otimes (\kappa_{\nu l}b_{\nu l}^{\dagger}+ \kappa^{*}_{\nu l}b_{\nu l})\right),\label{equ:H_SR2}\\
S_{L}=&a^{\dagger}+a,\ S_{R}=\rm\sin \theta \sigma_{z}+\rm\cos \theta \sigma_{x}.\label{jump operator2}
\end{align}
Considering the system internal interaction are strong even ultrastrong, while system-reservoir coupling are weak. Hence we can express the jump operators (see Eq. (\ref{jump operator2})) in the $H_{S}$ representation as
\begin{equation} \label{jump operator H_S}
S_{\nu}=\sum_{k} [\mathcal{S}_{\nu k}(\omega_{\nu k})+\mathcal{S}^{\dagger}_{\nu k}(\omega_{\nu k})].
\end{equation}
This equation including the positive and negative frequency components of the system operator $S_{\nu}$.
For simplification, we set $S_{\nu}^{+}=\sum_{k} \mathcal{S}_{\nu k}(\omega_{\nu k})$, and $S_{\nu}^{-}=\sum_{k} \mathcal{S}^{\dagger}_{\nu k}(\omega_{\nu k})$.
The Heisenberg equation of motion for the reservoir modes is
\begin{align} \label{b}
\frac{d b_{\nu l} (t)}{dt}=i [H (t), b_{\nu l} (t)]=-i \omega_{\nu l} b_{\nu l} (t)-i \kappa_{\nu l} S^{+}_{\nu}.
\end{align}
The Heisenberg equation of motion for SQUID and flux qubit is
\begin{align} \label{S}
\frac{d S^{+}_{\nu} (t)}{dt}=i [H (t), S^{+}_{\nu} (t)]=i [H_{S} (t), \mathcal{S}^{+}_{\nu} (t)]-i \sum_{l} \kappa^{*}_{\nu l} b_{\nu l} (t).
\end{align}
Since the evolution of reservoir modes $b_{\nu l} (t)$ are linear and hence Eq. (\ref{b}) can be solved as
\begin{align} \label{the solution of b}
b_{\nu l}(t)=b_{\nu l}(t_0) e^{-i \omega_{\nu l} (t-t_{0})}-i \kappa_{\nu l} \int^{t}_{t_0} dt^{\prime} S^{+}_{\nu} (t^{\prime}) e^{-i \omega_{\nu l} (t-t^{\prime})},
\end{align}
where $t_0<t$ denotes some initial time in the past.
We substitute Eq. (\ref{the solution of b}) into Eq. (\ref{S}) and obtain the following form
\begin{align} \label{dS} \nonumber
\frac{d {S}^{+}_{\nu} (t)}{dt} &=i [H_{S} (t), {S}^{+}_{\nu}]-\int^{t}_{t_0} d t^{\prime} \sum_{l} |\kappa_{\nu l}|^2 e^{-i \omega_{\nu l}(t-t^{\prime})} S^{+}_{\nu} (t^{\prime}) \\  &-\sum_{ l} i \kappa^{*}_{\nu l} e^{-i \omega_{\nu l} (t-t_0)} b_{\nu l} (t_0).
\end{align}
In order to solve this equation, we follow the method of Ref. \cite{girvin2011circuit} and we make the Markov approximation and consider the ohmic spectral function $J(\omega)=\sum_k\left|g_k\right|^2 \delta\left(\omega-\omega_k\right)$  with $g_{k} \propto \sqrt{\omega_{k}}$ \cite{weiss2012quantum, PhysRevA.104.053701},
\begin{align} \nonumber
&\sum_{l} |\kappa_{\nu l}|^2 e^{-i \omega_{\nu l}(t-t^{\prime})}  \\ \nonumber &  
=\int^{\infty}_{-\infty} d \omega \left[ \sum_l |\kappa_{\nu l}|^2 \delta\left(\omega-\omega_{\nu l}\right) \right] e^{-i \omega_{\nu l} (t-t^{\prime})} \\ &=\gamma \omega \delta(t-t^{\prime}).
\end{align}
Employing
$\int^{x_0}_{-\infty} \delta (x-x_0)=\frac{1}{2}$, the Eq.(\ref{dS}) can be resolved as
\begin{align}
\frac{d S^{+}_{\nu}}{dt} &=i [H_{S} (t), S^{+}_{\nu} (t)]-\frac{\gamma \omega}{2} S^{+}_{\nu}-\sum_{\nu l} i \kappa^{*}_{\nu l} e^{-i \omega_{\nu l} (t-t_0)} b_{\nu l} (t_0).
\end{align}
We now define the input field as
\begin{align} \label{b_in}
b_{in} =\frac{1}{\sqrt{2 \pi}}\sum_{l}  \sqrt{\omega_{\nu l}} e^{-i \omega_{\nu l} (t-t_0)} b_{\nu l} (t_0).
\end{align}
Hence
\begin{align}
\dot{S^{+}_{\nu}}=i [H_{S} (t), \mathcal{S}^{+}_{\nu} (t)]-\frac{\gamma \omega}{2} S^{+}_{\nu}-i \sqrt{\gamma} b_{in} (t).
\end{align}
To analyze a relation between output field and input field, one can employ an alternative solution to Eq. (\ref{b}) with $t_{1}>t$ for the future time,
\begin{align} \label{the solution of b in the future}
b_{\nu l}(t)=b_{\nu l}(t_1) e^{-i \omega_{\nu l} (t-t_{1})}-i \kappa^{*}_{\nu l} \int^{t_1}_{t} dt^{\prime} S^{+}_{\nu} (t^{\prime}) e^{-i \omega_{\nu l} (t-t^{\prime})}.
\end{align}
Defining the output field as
\begin{align} \label{b_out}
b_{out} (t)=\frac{1}{\sqrt{2 \pi}}\sum_{l}  \sqrt{\omega_{\nu l}} e^{-i \omega_{\nu l} (t-t_1)} b_{\nu l} (t_1).
\end{align}
We follow the same procedure as before and obtain 
\begin{align}
\dot{\mathcal{S}^{+}_{\nu}}=i [H_{S} (t), S^{+}_{\nu} (t)]+\frac{\gamma\omega}{2} \mathcal{S}^{+}_{\nu}-i \sqrt{\gamma} b_{out} (t).
\end{align}
Combining the Eq.(\ref{b_in}) and Eq. (\ref{b_out}), we can obtain the input-output relation as follows
\begin{align}
b_{out} (t)=b_{in} (t)-i \sqrt{\gamma} \omega S^{+}_{\nu}.
\end{align}
So this relation can also expressed as
\begin{equation}
b_{out} (t)=b_{in} (t)+ \sqrt{\gamma} \dot{S}_{\nu}^{+}.
\end{equation}

\bibliography{thermal_diode}
\end{document}